\documentclass[prd,aps,10pt,twocolumn,floatfix,nofootinbib,superscriptaddress,notitlepage,longbibliography]{revtex4-2}
\usepackage{amsmath,amssymb,bm}
\usepackage{needspace}
\usepackage{graphicx}
\usepackage{color}
\definecolor{revisionblue}{RGB}{0,0,0}
\newcommand{\DIFadd}[1]{{\color{revisionblue}#1}}

\usepackage{placeins}
\usepackage[T1]{fontenc}
\usepackage{hyperref}
\hypersetup{colorlinks=true,allcolors=black,
  pdftitle={Transferable neural-network preconditioners for Wilson-Clover propagator inversions},
  pdfauthor={Shiyi Zhong, Yuanyi Zhang, Jian-Hui Zhang, Dian-Jun Zhao}}
\DeclareRobustCommand{\revnov}[1]{#1}
\newcommand{\cfg}[2]{\ensuremath{\mathrm{cfg}#1_{\mathrm{C}#2}}}

\begin{document}
\title{Transferable neural-network preconditioners for Wilson--Clover propagator inversions}
\author{Shiyi Zhong}
\thanks{These authors contributed equally to this work.}
\affiliation{School of Science and Engineering, The Chinese University of Hong Kong (Shenzhen), Longgang, Shenzhen, Guangdong, 518172, P.R. China}

\author{Yuanyi Zhang}
\thanks{These authors contributed equally to this work.}
\affiliation{School of Science and Engineering, The Chinese University of Hong Kong (Shenzhen), Longgang, Shenzhen, Guangdong, 518172, P.R. China}

\author{Jian-Hui Zhang}
\email{Corresponding author: zhangjianhui@cuhk.edu.cn}

\affiliation{School of Science and Engineering, The Chinese University of Hong Kong (Shenzhen), Longgang, Shenzhen, Guangdong, 518172, P.R. China}

\affiliation{Southern Center for Nuclear-Science Theory (SCNT), Institute of Modern Physics, Chinese Academy of Sciences, Huizhou 516000, Guangdong Province, P.R. China}

\author{Dian-Jun Zhao}
\email{Corresponding author: zhaodianjun@cuhk.edu.cn}
\affiliation{School of Science and Engineering, The Chinese University of Hong Kong (Shenzhen), Longgang, Shenzhen, Guangdong, 518172, P.R. China}

\begin{abstract}
We investigate two single--layer neural-network architectures for accelerating four-dimensional Wilson--Clover propagator inversions on lattice ensembles in practical use; both transfer directly without retraining. The first is a gauge-covariant parallel transport convolution (PTC) network that directly reduces the number of generalized conjugate residual (GCR) solver iterations. The second is a CNN-like network used together with a multigrid-inspired algorithm. The two approaches can be combined to further reduce GCR iterations. PTC transfers without retraining across lattice volumes, lattice spacings, and pion masses while maintaining comparable acceleration, whereas the CNN-like network transfers directly from small- to large-volume ensembles with strong acceleration. A cost comparison accounting for both basis construction and solves shows that the proposed scheme is more advantageous for large-volume ensembles than for small-volume ones.
\end{abstract}
\maketitle
\raggedbottom

\section{Introduction}
\label{sec:intro}
Hadron spectroscopy and matrix-element calculations in lattice quantum chromodynamics (QCD) require many quark propagators on each gauge configuration. Each source entails a large Dirac-equation solve, while the gauge background and valence mass can remain fixed. Information prepared for one inversion can therefore help subsequent sources. The relevant computational cost includes both this preparation and the inversions required by the measurement.

Propagator solvers have a long history of exploiting the structure of the lattice Dirac operator. Even--odd preconditioning and nonsymmetric Krylov methods reduce the work of Wilson-fermion inversions~\cite{DeGrandRossi1990,Frommer1994}. Domain decomposition supplies local solves~\cite{Luscher2004}, while deflation and recycling reuse slowly converging directions~\cite{Luscher2007,StathopoulosOrginos2010}. The local coherence of low modes makes their representation on small lattice blocks economical. Adaptive multigrid combines a correction in this block representation with fine-lattice relaxation~\cite{Brannick2008,Babich2010}; combining multigrid with domain decomposition further improves Wilson--Clover solvers~\cite{Frommer2014}. \DIFadd{Recent studies have refined basis preparation through chiral truncation and local singular vectors~\cite{Whyte2025Truncation,Whyte2025SVD}; mixed-precision and multigrid implementations provide complementary advances~\cite{Clark2010,Clark2016}.}

Neural networks offer another way to construct an inexpensive approximate inverse. Ref.~\cite{LehnerWettig2023} introduced gauge-equivariant parallel-transport convolution (PTC) for lattice-QCD solvers, followed by gauge-equivariant pooling and unpooling for multigrid networks~\cite{LehnerWettigPooling2024}. Ref.~\cite{Pfahler2026} studied low-mode-sensitive training and transfer between gauge backgrounds and volumes. \DIFadd{Ref.~\cite{ChoiOhno2026} investigated learned low-mode test vectors for multigrid setup.} Learned preconditioners have also been developed for the normal equations of the two-dimensional Schwinger model~\cite{Cali2023,Sun2026}. \DIFadd{These studies motivate our investigation of local learned corrections in four-dimensional Wilson--Clover propagator solves.}

\DIFadd{We investigate whether single-layer local maps, whose parameter counts do not grow with lattice volume, can reduce the outer generalized conjugate residual (GCR) iterations required for Wilson--Clover propagator solves, and whether fixed trained weights remain effective across gauge configurations and ensembles. We assess each solve by convergence of the true relative residual to the prescribed tolerance and by its outer GCR iteration count. Dirac-operator application counts provide a separate estimate of computational work; for multiple solves, we include the stated basis-preparation costs and treat the trained network weights as available.}

\DIFadd{We study two operators. The gauge-covariant PTC is a single fine-lattice layer using an on-site path and eight nearest-neighbor paths. The other operator resembles a convolutional neural network (CNN): it applies a shared linear map to the coefficients of the current block and its eight neighboring blocks in a local basis rebuilt for each target gauge field. Here CNN-like refers to shared local weights in a single linear layer, with no nonlinear activation. Within the multigrid-inspired correction, it replaces the reference projection onto global templates, after which the same relaxation is applied. We test PTC on its own, fit it directly to near-null error, and combine it with the multigrid-inspired correction at three different stages. Both maps reduce GCR iterations after transfer to other ensembles, including larger lattice volumes. Adding PTC to the CNN-like correction further reduces outer GCR iterations in the matched tests on CLQCD ensembles.}

The rest of the paper is organized as follows. Sec.~\ref{sec:theory} introduces the PTC and CNN-like corrections within the propagator solver. Sec.~\ref{sec:numerics} evaluates their transfer, composition, and preparation-plus-solve costs. Sec.~\ref{sec:conclusion} contains our conclusions. Appendices~\ref{app:sec:theory}--\ref{app:sec:training} provide a self-contained account of the basis construction, projected reference solve, training definitions, and GCR iteration.

\section{Theoretical analysis}
\label{sec:theory}

We construct two learned local corrections within a conventional Krylov solver for quark propagators. The reference combines a projected solve in a small set of template fields with fine-lattice relaxation, following deflation and multigrid methods~\cite{Luscher2007,Brannick2008,Babich2010,Frommer2014}. PTC acts on fine-lattice quark fields; the CNN-like map replaces the template correction by a shared local operation on block coefficients. We give the operations needed to interpret the numerical comparisons below. The full basis construction and GCR recurrence are given in Appendices~\ref{app:subsec:preparation} and \ref{app:subsec:source-loop}, with notation defined independently of the main text.

For a fixed gauge field $U$, dimensionless bare valence mass $m_0$, and clover coefficient $c_{\rm SW}$, let $D=D[U,m_0]$ denote the Wilson--Clover Dirac operator~\cite{SheikholeslamiWohlert1985}, with its dependence on $c_{\rm SW}$ implicit. A lattice with extents $L_x,L_y,L_z,L_t$ in lattice units has $V=L_xL_yL_zL_t$ sites. The quark fields have $n=4\times3\times V$ complex components, accounting for spin, color, and position. The source $\eta\in\mathbb C^n$ and propagator column $\psi\in\mathbb C^n$, where $\mathbb C$ denotes the complex numbers, satisfy
\begin{equation}
 D\psi=\eta.
 \label{eq:solve}
\end{equation}
At the GCR iteration $j$, the field $\psi_j$ has an unknown error $e_j=\psi-\psi_j$ satisfying
\begin{equation}
 De_j=\eta-D\psi_j=r_j,
 \label{eq:error-equation}
\end{equation}
where $r_j$ is the residual. A preconditioner supplies a trial correction $t_j\in\mathbb C^n$ that approximates this error. GCR uses $t_j$ and its Dirac image $Dt_j$ to construct a search direction and update $\psi_j$ by minimizing the original residual~\cite{Eisenstat1983}. Both learned maps remain fixed during a solve. We use $\|\cdot\|_2$ for the Euclidean norm over all field components, $\dagger$ for the Hermitian adjoint, and $I_N$ for the identity on an $N$-dimensional space.

\subsection{PTC on the quark-field residual}
\label{subsec:ptc-learning}

The gauge-covariant PTC layer~\cite{LehnerWettig2023,LehnerWettigPooling2024} uses one on-site path and eight nearest-neighbor paths. Let $v\in\mathbb C^n$ be an input quark field and $x$ a fine-lattice site. The displacement $\delta$ is zero or one lattice spacing in either direction along a lattice axis. The color transporter $\mathcal U_\delta(x)$ brings $v(x+\delta)$ to $v(x)$, and a learned matrix $A_\delta\in\mathbb C^{4\times4}$ mixes its spin components. The map $F_{\rm PTC}$ is
\begin{equation}
 (F_{\rm PTC}v)(x)=\sum_\delta
 [A_\delta\otimes\mathcal U_\delta(x)]v(x+\delta).
 \label{eq:ptc}
\end{equation}
Here $\otimes$ denotes the direct product of the spin and color. The transporters are $3\times3$ matrices in the special unitary color group $\mathrm{SU}(3)$, constructed from stout-smeared links~\cite{Morningstar2004}; backward paths use adjoint links, and the on-site transporter is $I_3$. Parallel transport ensures that $F_{\rm PTC}v$ transforms as a quark field.

All sites share the nine spin matrices $A_\delta$, giving $9\times4^2=144$ complex parameters. Training uses $N_p$ random quark fields $v_\ell$, with probe index $\ell=1,\ldots,N_p$. The training input is $Dv_\ell$ and the target is $v_\ell$, so the training loss is
\begin{equation}
 \mathcal L_{\rm PTC}=\frac{1}{N_p}\sum_{\ell=1}^{N_p}
 \frac{\|F_{\rm PTC}Dv_\ell-v_\ell\|_2^2}{\|v_\ell\|_2^2}.
 \label{eq:ptc-loss}
\end{equation}
After training, the same spin matrices can be reused on another configuration.

PTC supplies $t_j=F_{\rm PTC}r_j$.
\revnov{Let $B:\mathbb C^n\to\mathbb C^n$ denote the complete baseline correction, including any relaxation. We investigate three variants of PTC. ``PTC first'' applies $F_{\rm PTC}$ before $B$:
\begin{equation}
 t_j=B(F_{\rm PTC}r_j).
 \label{eq:pre}
\end{equation}
``PTC last'' applies $F_{\rm PTC}$ after $B$:
\begin{equation}
 t_j=F_{\rm PTC}(Br_j).
 \label{eq:post}
\end{equation}
``PTC residual'' applies $F_{\rm PTC}$ to the residual left by the complete correction $Br_j$:
\begin{equation}
 t_j=Br_j+F_{\rm PTC}\bigl(r_j-D(Br_j)\bigr).
 \label{eq:residual-insertion}
\end{equation}
Note that ``PTC residual'' adds one application of $D$ per preconditioner call, while ``PTC first'' and ``PTC last'' do not.} In the construction of $B$ below, its input is denoted by $q\in\mathbb C^n$:
\begin{equation}
 q=
 \begin{cases}
 r_j, & \text{without PTC preprocessing},\\
 F_{\rm PTC}r_j, & \text{with PTC preprocessing}.
 \end{cases}
 \label{app:eq:pre-input}
\end{equation}

\subsection{CNN-like correction in the local block basis}
\label{subsec:coarse-learning}

The reference represents a correction with a small set of full-lattice templates. We first prepare $N_v=32$ candidate quark fields. Starting from random fields $v_i^{(0)}$, with candidate index $i=1,\ldots,N_v$, $k$ Richardson steps give
\begin{equation}
 v_i^{(k)}=(I_n-\omega_0D)^k v_i^{(0)},
 \label{eq:candidates}
\end{equation}
where $\omega_0$ is the fixed preparation step size. Alternatively, Lanczos iteration prepares the eigenvectors of $D^\dagger D$ with the smallest eigenvalues, namely the right singular vectors of $D$ with the smallest singular values. Either construction gives a candidate matrix $\mathcal V\in\mathbb C^{n\times N_v}$ whose columns are complete candidate quark fields. The preparation length $k$ is separately chosen from the number of relaxation steps (which is taken as 5 in the present case) used in each preconditioner call.

Divide the lattice into $N_b=V/16$ blocks, with each of them having $2^4=16$ lattice sites and $4\times3\times16=192$ spin--color components, where the latter are orthonormalized by a QR factorization. Here QR denotes a factorization into orthonormal columns and an upper-triangular matrix. We retain $s=8$ local vectors per block, giving $n_c=N_bs$ block coefficients. The subscript $c$ labels the block representation, not color. The precise column selection is given in Appendix~\ref{app:subsec:local-basis}.

Extend each local vector by zero outside its block and collect them in $P\in\mathbb C^{n\times n_c}$. Then $P$ reconstructs a quark field from block coefficients, while $R=P^\dagger\in\mathbb C^{n_c\times n}$ extracts those coefficients by projection. Local orthonormality gives $P^\dagger P=I_{n_c}$. These maps are rebuilt on every target gauge configuration.

The reference further compresses the block representation. \revnov{Let $N_d\leq N_v$ be the numerical rank of $R\mathcal V$. The matrix $W\in\mathbb C^{n_c\times N_d}$ contains an orthonormal basis for its retained column space:}
\begin{equation}
 W=\operatorname{orth}(R\mathcal V).
 \label{eq:global-basis}
\end{equation}
\revnov{Here $\operatorname{orth}$ discards numerically dependent directions and orthonormalizes the retained columns.} Reconstructing these columns gives the template matrix
\begin{equation}
 Z=PW.
 \label{eq:selected-fields}
\end{equation}
Thus $Z\in\mathbb C^{n\times N_d}$ contains $N_d$ complete candidate quark fields, each with $n$ components. The projected Dirac matrix is
\begin{equation}
 H=Z^\dagger DZ\in\mathbb C^{N_d\times N_d}.
 \label{eq:retained-dirac}
\end{equation}
For input $q$, the vector of global template weights $\xi\in\mathbb C^{N_d}$ is found by a direct solve of
\begin{equation}
 (H+\tau I_{N_d})\xi=Z^\dagger q,
 \label{eq:small-solve}
\end{equation}
where $\tau\geq0$ is a prescribed real shift to regularize potential singularities in $H$. The first quark-field correction is then
\begin{equation}
 \delta\psi_c=Z\xi.
 \label{eq:reference-correction}
\end{equation}
The subscript $c$ identifies the contribution from the block representation. The shifted matrix is factored once per target configuration. For $N_d=32$, every call solves a $32\times32$ system, but its correction is restricted to 32 global template weights.

The CNN-like map replaces this first correction. Its input is the projected field $Rq\in\mathbb C^{n_c}$, with $s$ coefficients $(Rq)(y)$ on block $y$. The displacement $\Delta$ is zero or one block in either direction along a lattice axis. Nine shared matrices $T_\Delta\in\mathbb C^{s\times s}$ define $F_{\rm CNN}:\mathbb C^{n_c}\to\mathbb C^{n_c}$ by
\begin{equation}
 (F_{\rm CNN}Rq)(y)=\sum_\Delta T_\Delta(Rq)(y+\Delta).
 \label{eq:coarse-stencil}
\end{equation}
The block shifts are periodic. Each block combines its own residual coefficients with those of its eight neighbors to produce eight correction coefficients. This single linear layer has no nonlinear activation and uses $9s^2=9\times8^2=576$ complex parameters, shared by all blocks. Its output has $n_c$ coefficients even though its parameter count is independent of volume. Expanding back to the fine lattice gives
\begin{equation}
 \delta\psi_c=PF_{\rm CNN}(Rq).
 \label{eq:learned-correction}
\end{equation}
Equation~\eqref{eq:learned-correction} replaces Eqs.~\eqref{eq:small-solve} and \eqref{eq:reference-correction}; applying $F_{\rm CNN}$ requires neither a projected solve nor the global templates $W,Z$.

\revnov{After either first correction, the remaining input is}
\begin{equation}
 r_c=q-D\delta\psi_c.
 \label{eq:coarse-residual}
\end{equation}
Five Richardson steps approximately solve for its remaining correction. With $u^{(0)}=0$, define the quark fields $u^{(h)}\in\mathbb C^n$ by
\begin{equation}
 u^{(h+1)}=u^{(h)}+\omega\bigl(r_c-Du^{(h)}\bigr),
 \label{eq:relaxation}
\end{equation}
where $h=0,\ldots,4$ counts these steps and $\omega$ is their fixed step size. Write $Sr_c=u^{(5)}$ for the resulting relaxation map. The complete baseline returns
\begin{equation}
 Bq=\delta\psi_c+Sr_c.
 \label{eq:cnn-baseline}
\end{equation}
The CNN-like variant changes only the first correction: it replaces Eqs.~\eqref{eq:small-solve} and \eqref{eq:reference-correction} by Eq.~\eqref{eq:learned-correction}, then uses the same residual update and five Richardson steps. Without a first correction, the relaxation-only baseline is $B=S$. GCR receives $t_j=Bq$, or the corresponding combined correction from Sec.~\ref{subsec:ptc-learning}.

\begin{figure*}[!tp]
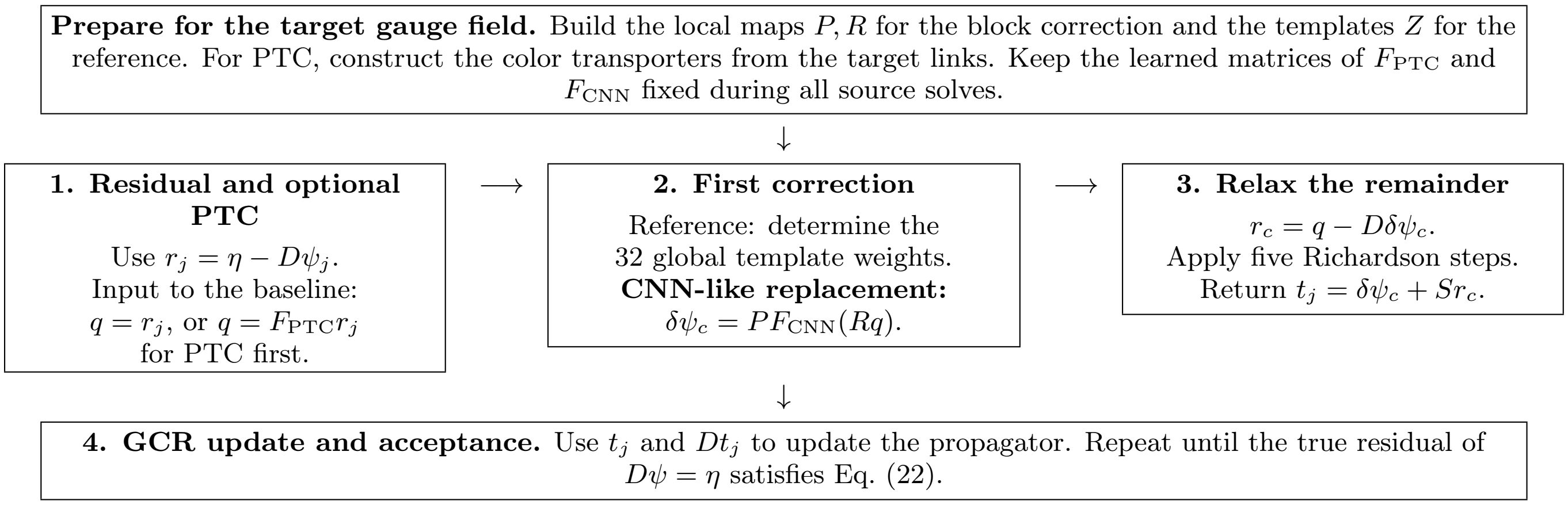

\centering
\fbox{\begin{minipage}{0.94\textwidth}
\small\textbf{Prepare for the target gauge field.}
Build the local maps $P,R$ for the block correction and the templates $Z$ for the reference.
For PTC, construct the color transporters from the target links.
Keep the learned matrices of $F_{\rm PTC}$ and $F_{\rm CNN}$ fixed during all source solves.
\end{minipage}}
\par\smallskip$\downarrow$\par\smallskip
\fbox{\begin{minipage}[t]{0.27\textwidth}
\small\textbf{1. Residual and optional PTC}\\[3pt]
Use $r_j=\eta-D\psi_j$.\\
Input to the baseline:\\
$q=r_j$, or $q=F_{\rm PTC}r_j$\\
for PTC first.
\end{minipage}}
\hfill$\longrightarrow$\hfill
\fbox{\begin{minipage}[t]{0.29\textwidth}
\small\textbf{2. First correction}\\[3pt]
Reference: determine the\\
32 global template weights.\\
\textbf{CNN-like replacement:}\\
$\delta\psi_c=PF_{\rm CNN}(Rq)$.
\end{minipage}}
\hfill$\longrightarrow$\hfill
\fbox{\begin{minipage}[t]{0.27\textwidth}
\small\textbf{3. Relax the remainder}\\[3pt]
$r_c=q-D\delta\psi_c$.\\
Apply five Richardson steps.\\
Return $t_j=\delta\psi_c+Sr_c$.
\end{minipage}}
\par\smallskip$\downarrow$\par\smallskip
\fbox{\begin{minipage}{0.94\textwidth}
\small\textbf{4. GCR update and acceptance.}
Use $t_j$ and $Dt_j$ to update the propagator.
Repeat until the true residual of $D\psi=\eta$ satisfies Eq.~\eqref{eq:acceptance}.
\end{minipage}}
\caption{Placement of the learned corrections in a propagator solve. The CNN-like map replaces the first template correction; PTC first changes the input to the complete baseline. PTC alone bypasses the block correction and relaxation and returns $F_{\rm PTC}r_j$ directly to GCR. The detailed preparation, projected reference solve, and GCR updates are given in Appendices~\ref{app:subsec:preparation} and \ref{app:subsec:source-loop}.}
\label{fig:workflow}
\end{figure*}

Training determines $F_{\rm CNN}$ before the propagator solves. For each residual $q$ recorded in an earlier GCR iteration, we obtain target block-correction coefficients $\xi_c\in\mathbb C^{n_c}$ by approximately solving
\begin{equation}
 (RDP)\xi_c=Rq.
 \label{eq:coarse-system}
\end{equation}
Equation~\eqref{eq:coarse-system} uses the same projected Dirac construction as Eqs.~\eqref{eq:retained-dirac}--\eqref{eq:reference-correction}, but in the full block basis $P$ rather than the compressed template basis $Z$, to generate training targets for $F_{\rm CNN}$. Thus $RDP$ acts on $n_c$ block coefficients, whereas $H$ acts on $N_d$ global weights. The target fine-lattice correction is $P\xi_c$; during a propagator solve, the network supplies the coefficients $F_{\rm CNN}(Rq)$ instead. For 64 training pairs $(Rq_\ell,\xi_{c,\ell})$, indexed by $\ell$, we minimize
\begin{equation}
 \begin{aligned}
 \mathcal L_{\rm CNN}={}&\sum_{\ell=1}^{64}\|F_{\rm CNN}(Rq_\ell)-\xi_{c,\ell}\|_2^2\\
 &+\lambda\sum_{\Delta}\|T_\Delta\|_F^2.
 \end{aligned}
 \label{eq:coarse-loss}
\end{equation}
Here $\lambda\geq0$ controls the penalty on large matrix entries, and the Frobenius norm $\|\cdot\|_F$ is the square root of the sum of squared entry magnitudes. Training and transfer use the same local-basis convention. The fitted matrices remain fixed while $P$ and $R$ are rebuilt for the target gauge field.

Figure~\ref{fig:workflow} summarizes the complete solve. The learned weights can be shared across sources and volumes; target links and local bases supply the gauge-field dependence. For every method, GCR recomputes the true residual of Eq.~\eqref{eq:solve} and accepts convergence only when
\begin{equation}
 \frac{\|\eta-D\psi_j\|_2}{\|\eta\|_2}\leq10^{-8}.
 \label{eq:acceptance}
\end{equation}
Learning changes the trial correction, while the original Dirac equation fixes the accepted propagator.

\section{Numerical Setup and Analysis}
\label{sec:numerics}

\subsection{Gauge ensembles, templates, and training}
\label{subsec:setup}
Table~\ref{tab:ensembles} lists the seven ensembles used in the numerical analysis. C24P29 and F32P30 are Chinese Lattice QCD (CLQCD) ensembles reported in Ref.~\cite{Hu2024CLQCD}; C12P29 is an additional, previously unpublished CLQCD ensemble. The four A ensembles are pure-gauge ensembles generated for the transfer tests.

PTC weights are trained on the three CLQCD ensembles and compared with models trained locally on each A ensemble. The block-basis and composition tests use C12P29 and C24P29; valence parameters are specified for each test series.

\begin{table}[htbp]
\caption{Ensembles used in this work. TITLC and TITLS denote the tadpole-improved tree-level clover fermion and Symanzik gauge actions, respectively~\cite{Hu2024CLQCD}. The volume $V$ counts lattice sites. The gauge coupling parameter $\beta$ follows the convention of that reference for the CLQCD ensembles; values in this table are displayed to one decimal place.}
\label{tab:ensembles}
\small
\setlength{\tabcolsep}{3pt}
\begin{ruledtabular}
\begin{tabular}{llcc}
Ensemble & Ensemble type & Volume $V$ & $\beta$\\
\hline
C12P29 & TITLC + TITLS & $12^3\times32$ & 6.2\\
C24P29 & TITLC + TITLS & $24^3\times72$ & 6.2\\
F32P30 & TITLC + TITLS & $32^3\times96$ & 6.4\\
$\mathrm{A}_{\beta\,6.2}8$ & TITLC + TITLS & $8^3\times16$ & 6.2\\
$\mathrm{A}_{\beta\,6.4}8$ & TITLC + TITLS & $8^3\times16$ & 6.4\\
$\mathrm{A}_{\beta\,6.6}8$ & TITLC + TITLS & $8^3\times16$ & 6.6\\
$\mathrm{A}_{\beta\,6.8}8$ & TITLC + TITLS & $8^3\times16$ & 6.8
\end{tabular}
\end{ruledtabular}
\end{table}

\revnov{CLQCD configurations are labeled $\mathrm{cfg}i_{\mathrm{C}12}$, $\mathrm{cfg}i_{\mathrm{C}24}$, or $\mathrm{cfg}i_{\mathrm{F}32}$, where $i$ is the configuration number. For the A ensembles, $\mathrm{cfg}i_{\mathrm{b}2}$, $\mathrm{cfg}i_{\mathrm{b}4}$, $\mathrm{cfg}i_{\mathrm{b}6}$, and $\mathrm{cfg}i_{\mathrm{b}8}$ denote $\beta=6.2,6.4,6.6,6.8$, respectively. The block-basis and composition tests use $c_{\rm SW}=1.1609$ and $m_0=-0.286$, with additional C24P29 tests at $m_0=-0.277$. Standalone PTC transfer on C24P29 uses $m_0=-0.280$ and $c_{\rm SW}=1.16$. Comparisons use matched parameters within each series.}

\revnov{We prepare 32 candidates using Eq.~\eqref{eq:candidates} or Lanczos iteration and normally retain 32 global templates through Eq.~\eqref{eq:global-basis}. Preparation and the five-step relaxation in Eq.~\eqref{eq:relaxation} use the same step size. The reference solve follows Eq.~\eqref{eq:small-solve}. GCR starts from zero, uses restart length 50 and two orthogonalization passes, and tests convergence with Eq.~\eqref{eq:acceptance}.}

\revnov{PTC uses one stout-smearing step with smearing parameter $\rho=0.125$~\cite{Morningstar2004}. We train the layer in Eq.~\eqref{eq:ptc} with the loss in Eq.~\eqref{eq:ptc-loss}, using adaptive moment estimation (Adam) at learning rate $10^{-3}$ for 900 steps. The transferred models are trained on \cfg{2000}{12}, \cfg{4050}{24}, and $\mathrm{cfg}1000_{\mathrm{F}32}$. The first two use $m_0=-0.286$ and $c_{\rm SW}=1.16$; the third uses $m_0=-0.275$ and $c_{\rm SW}=0.956955$. Local controls use the same optimization settings on $\mathrm{cfg}120_{\mathrm{b}2}$, $\mathrm{cfg}120_{\mathrm{b}4}$, $\mathrm{cfg}120_{\mathrm{b}6}$, and $\mathrm{cfg}120_{\mathrm{b}8}$. Each control is trained at a mass different from both test masses on its ensemble, and its training configuration is excluded from the test set.}

\revnov{The CNN-like map is trained on \cfg{2000}{12} at $m_0=-0.286$ using 64 pairs from earlier GCR solves. Targets are approximate solutions of Eq.~\eqref{eq:coarse-system}, and the fit minimizes Eq.~\eqref{eq:coarse-loss}. Both learned maps remain fixed during subsequent solves and transfer tests.}

\needspace{6\baselineskip}
\subsection{Measuring work per propagator and across sources}
\label{subsec:cost-model}
\revnov{For the block-basis and composition tests, we estimate work by counting Dirac applications. Let $J$ count outer GCR iterations and $c_B$ count Dirac applications per preconditioner call. At restart length 50, the estimate per solve is
}\begin{equation}
 \widehat N_D(J,c_B)=J(c_B+1)+\left\lceil\frac{J}{50}\right\rceil+2.
 \label{eq:count}
\end{equation}\revnov{
The ceiling $\lceil\cdot\rceil$ rounds upward. The count includes preconditioner and GCR Dirac applications and residual checks; it excludes local-map arithmetic, projections, and communication.}

\revnov{Equations~\eqref{eq:coarse-residual} and \eqref{eq:relaxation} give $c_B=5$ for either the reference or CNN-like correction and $c_B=4$ for relaxation alone. Standalone PTC [Eq.~\eqref{eq:ptc}] has $c_B=0$. PTC first and last [Eqs.~\eqref{eq:pre} and \eqref{eq:post}] preserve these counts; PTC residual [Eq.~\eqref{eq:residual-insertion}] adds one Dirac application per call.}

\revnov{Let $A_D$ be the preparation cost and $J_\nu$ the iteration count for source $\nu$. For $N_{\rm src}$ source solves, the total estimated work is
}\begin{equation}
 T_D(N_{\rm src})=A_D+\sum_{\nu=1}^{N_{\rm src}}\widehat N_D(J_\nu,c_B).
 \label{eq:budget}
\end{equation}\revnov{
Preparing relaxed candidates [Eq.~\eqref{eq:candidates}] costs $N_vk$ Dirac applications; assembling the projected matrix [Eq.~\eqref{eq:retained-dirac}] costs $N_d$. Each Lanczos application of $D^\dagger D$ counts as two. We charge the recorded basis preparation and matrix assembly, assuming trained weights are available; training and gauge preprocessing are excluded. Four-source work estimates are averaged after applying Eq.~\eqref{eq:count} to each source.}

Iteration comparisons use the reduction
\begin{equation}
 \mathcal R_{\rm GCR}=100\left(1-\frac{J_{\rm method}}{J_{\rm baseline}}\right)\%.
 \label{eq:iteration-reduction}
\end{equation}
Here $J_{\rm method}$ and $J_{\rm baseline}$ are the outer GCR iteration counts under matched configuration, source, valence-mass, and convergence conditions. Positive values mean fewer iterations; negative values mean more. Each caption states the baseline and, where applicable, the averaging procedure.

\begin{figure*}[!tp]
\centering
\includegraphics[width=0.92\textwidth]{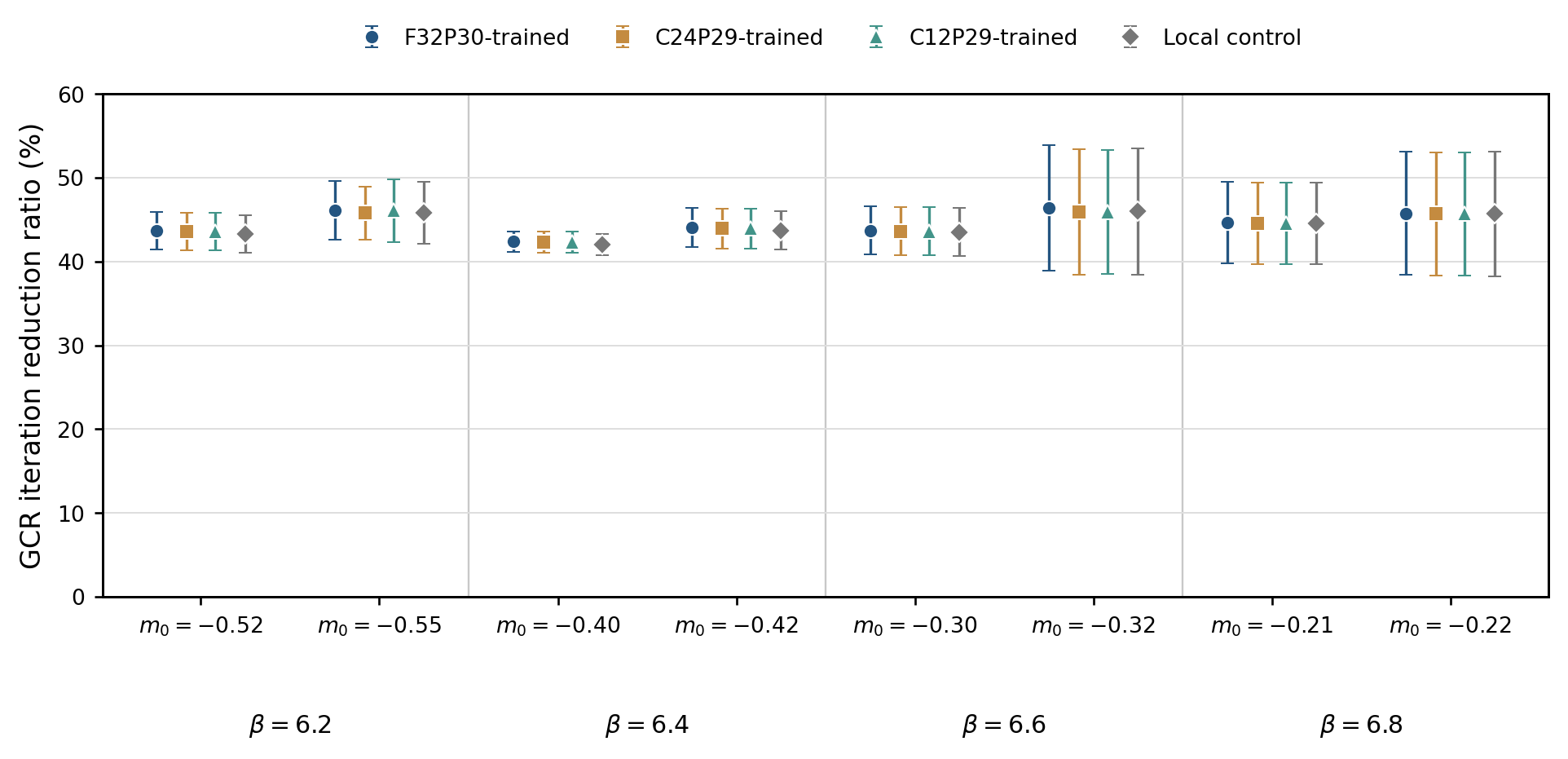}
\caption{GCR iteration reduction ratio, defined in Eq.~\eqref{eq:iteration-reduction}, for the single-layer PTC tests on the A ensembles listed in Table~\ref{tab:ensembles}. The tested methods are GCR with each of the three transferred PTC models or the local control; the baseline is unpreconditioned GCR on the same configuration and source at the same valence mass. Tick labels give $m_0$, grouped by $\beta$. Points show means over configurations, and error bars show one sample standard deviation of the per-configuration GCR iteration reduction ratios. Each point summarizes approximately 15 configurations. Colors distinguish the models. All included solves reach a true relative residual of $10^{-8}$; each point uses the configurations for which all five methods converged.}
\label{fig:fine-transfer}
\end{figure*}

\subsection{Transfer of a single PTC layer}
\label{subsec:fine-transfer}
\revnov{We transfer the PTC models trained on F32P30, C24P29, and C12P29 to the four A ensembles, using Eq.~\eqref{eq:ptc}. Each solve uses a point source at the origin in spin 0 and color 0, $c_{\rm SW}=1.16$, and the stout-smearing settings of Sec.~\ref{subsec:setup}. In Fig.~\ref{fig:fine-transfer}, transferred weights and local controls give similar iteration reductions across the tested $\beta$ and $m_0$ values. Reductions generally grow toward criticality within each mass pair.}

\revnov{On five held-out C24P29 configurations, we solve all twelve spin--color point-source columns at $m_0=-0.280$. PTC weights trained on \cfg{2000}{12} and \cfg{4050}{24}, both at $m_0=-0.286$, reduce the mean iteration counts by 31.5\% and 31.4\%, respectively, relative to unpreconditioned GCR. Other settings follow Sec.~\ref{subsec:setup}.}

\revnov{These tests show transfer without retraining across the investigated gauge backgrounds, lattice spacings, valence masses, and volumes.}

\subsection{Direct single-layer PTC fits to low-mode error}
\label{subsec:direct-low}
\revnov{We also fit PTC directly to low-mode error, without a block basis. For held-out errors projected onto the 32 lowest measured modes on C12P29, path radii of one to three lattice spacings leave a median error fraction of $0.99968$--$0.99987$. A matched flexible generalized minimal residual (FGMRES) solve~\cite{Saad1993} takes 8400 iterations with direct low-mode fitting and 8500 without it. This small gain motivates combining PTC with the block correction in Eq.~\eqref{eq:cnn-baseline}.}

\begin{figure}[!tbp]
\centering
\includegraphics[width=\linewidth]{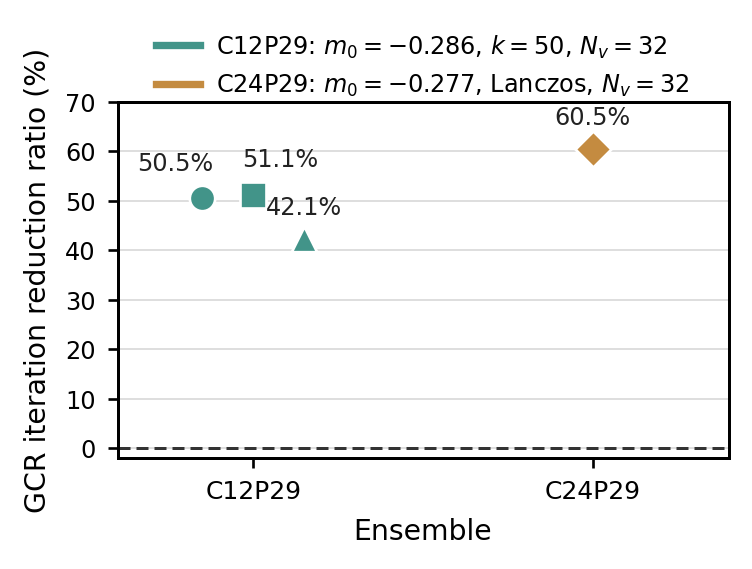}
\caption{GCR iteration reduction ratio, defined in Eq.~\eqref{eq:iteration-reduction}, for the CNN-like correction. The tested method uses the CNN-like correction followed by five Richardson relaxation steps; the baseline uses the same five relaxation steps alone. The dashed line denotes zero GCR iteration reduction ratio. Marker shapes identify configurations: circle, \cfg{2000}{12}; square, \cfg{2020}{12}; triangle, \cfg{3000}{12}; diamond, \cfg{4050}{24}. C12P29 uses target-specific $N_v=32$ block bases generated with $k=50$ Richardson steps; C24P29 uses 32 target-specific Lanczos modes. Each point is a single-configuration result, and all solves reach a true relative residual of $10^{-8}$. For the additional comparisons quoted in the text, \revnov{the baselines are the matched reference correction with a direct solve of the shifted template system and five Richardson steps on C24P29, and unpreconditioned GCR on C12P29.}}
\label{fig:cnn-transfer}
\end{figure}

\subsection{CNN-like correction and transfer}
\label{subsec:replacement}
\revnov{We test the CNN-like correction [Eqs.~\eqref{eq:learned-correction} and \eqref{eq:cnn-baseline}] with fixed C12P29-trained weights on C12P29 and C24P29. The C24P29 volume is 18 times that of C12P29. Each target basis is rebuilt from 32 random candidates after $k=50$ Richardson steps [Eq.~\eqref{eq:candidates}] on C12P29, or from 32 Lanczos vectors on C24P29. All solves in Fig.~\ref{fig:cnn-transfer} satisfy Eq.~\eqref{eq:acceptance}.}

\revnov{Relative to five-step relaxation alone [Eq.~\eqref{eq:relaxation}], the CNN-like correction reduces GCR iterations by 42.1\%--51.1\% in the three C12P29 tests and by 60.5\% on C24P29. Matched comparisons also give a 25.2\% reduction against the reference with relaxation [Eqs.~\eqref{eq:reference-correction} and \eqref{eq:cnn-baseline}] on C24P29, and reductions exceeding 90\% against unpreconditioned GCR on C12P29.}

\begin{figure}[!tbp]
 \centering
\includegraphics[width=\columnwidth]{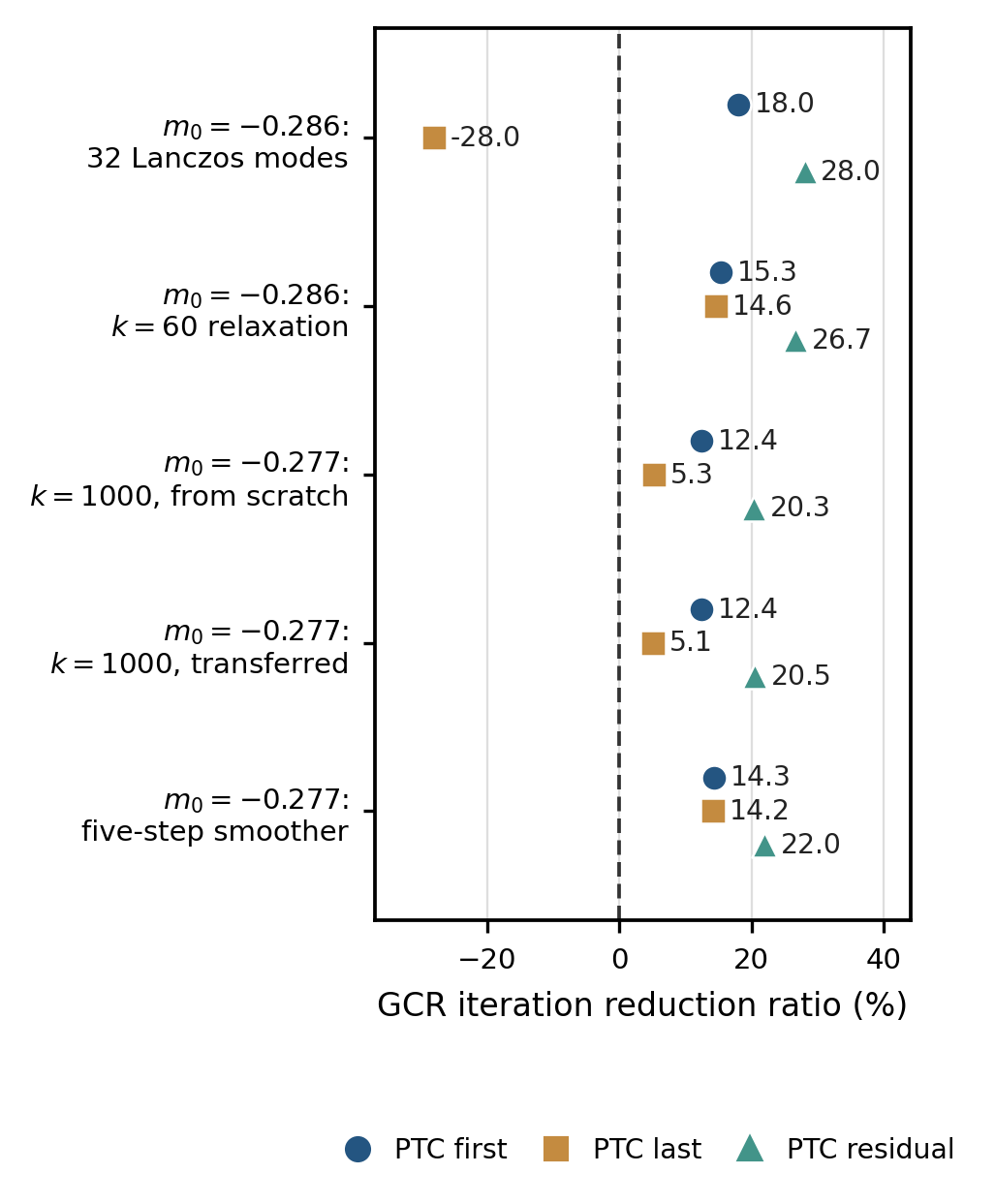}
\caption{GCR iteration reduction ratio, defined in Eq.~\eqref{eq:iteration-reduction}, for three PTC placements on C24P29. Each tested method combines PTC with the baseline correction in the indicated placement; its baseline omits PTC. In the first four rows, the baseline uses the reference template correction followed by five Richardson relaxation steps; in the last row, it uses five-step relaxation alone. Row labels specify the mass and basis. The first two rows use \cfg{4500}{24}, and the remaining rows use \cfg{4050}{24}. Circles, squares, and triangles denote PTC first, PTC last, and PTC residual. Each point is a matched single-source result; all solves reach a true relative residual of $10^{-8}$.}
\label{fig:ptc-composition-c24}
\end{figure}

\subsection{PTC placement in the combined correction}
\label{subsec:composition}
\begin{figure*}[!tp]
 \centering
\includegraphics[width=0.98\textwidth]{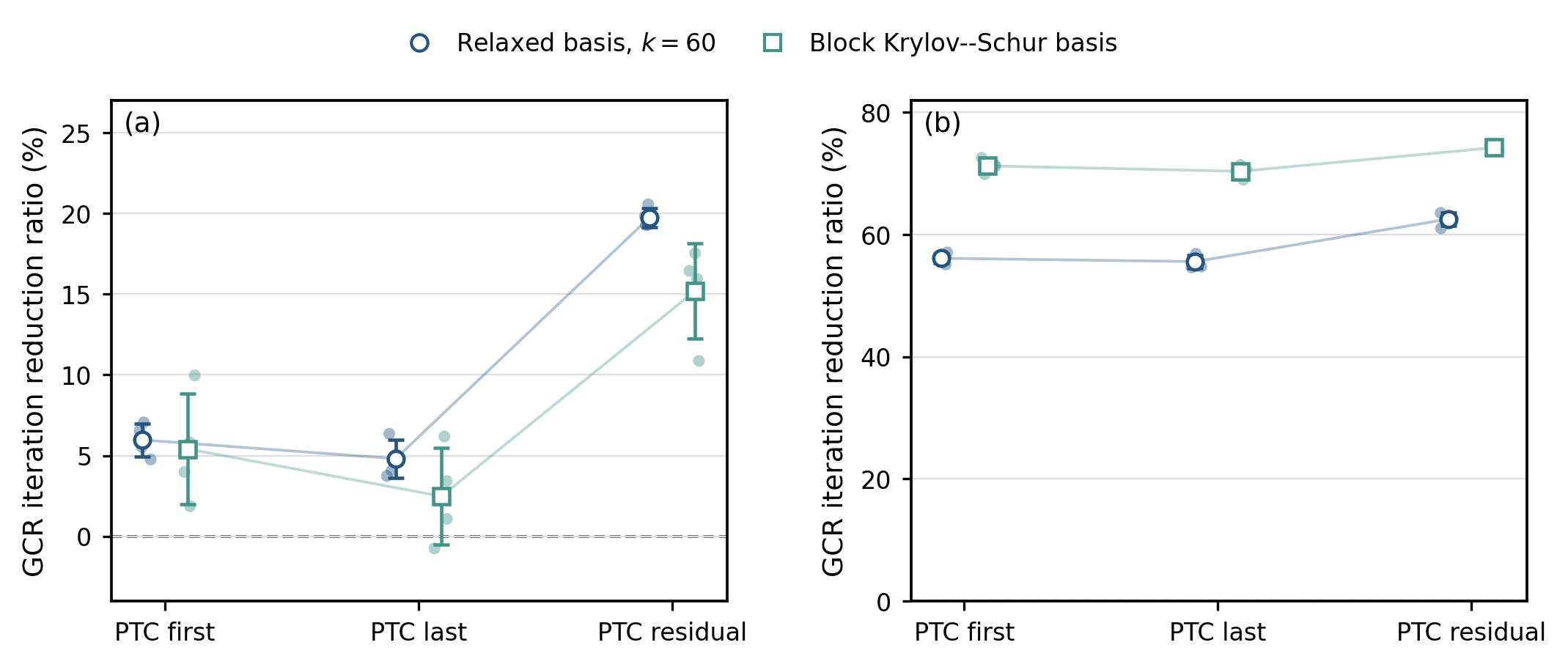}
\caption{GCR iteration reduction ratio, defined in Eq.~\eqref{eq:iteration-reduction}, for three PTC placements combined with the CNN-like correction on \cfg{2000}{12} from C12P29. In both panels, the tested method is the complete PTC--CNN-like correction, including five Richardson relaxation steps. The baseline in panel (a) is the corresponding CNN-like correction with relaxation but without PTC; the baseline in panel (b) is five-step relaxation alone. Circles and squares denote the $k=60$ relaxed basis and the block KS basis, respectively. Faint points show individual right-hand sides; open markers show the mean GCR iteration reduction ratios over four right-hand sides, with error bars denoting one sample standard deviation. All solves reach a true relative residual of $10^{-8}$.}
\label{fig:ptc-composition-c12}
\end{figure*}

\revnov{We compare PTC first, last, and residual [Eqs.~\eqref{eq:pre}, \eqref{eq:post}, and \eqref{eq:residual-insertion}] with the gauge field, source, valence mass, target basis, and baseline fixed within each comparison. Figures~\ref{fig:ptc-composition-c24} and \ref{fig:ptc-composition-c12} specify the baselines.}

\revnov{On C24P29, Fig.~\ref{fig:ptc-composition-c24} uses the reference and relaxation-only baselines, without the CNN-like map. PTC residual gives the largest iteration reductions in all five cases, 20.3\%--28.0\%, followed by PTC first at 12.4\%--18.0\%. PTC last increases iterations by 28.0\% in the Lanczos case and reduces them by 5.1\%--14.6\% in the other four cases.}

\revnov{Figure~\ref{fig:ptc-composition-c12} compares the PTC--CNN-like combinations on C12P29. Target bases use either 32 random candidates after $k=60$ Richardson steps [Eq.~\eqref{eq:candidates}] or thick-restart block Krylov--Schur (block KS) iteration with eight starting fields and 32 retained Ritz vectors. Averaged over four sources, the reductions relative to the CNN-like correction with relaxation are 15.2\%--19.7\% for PTC residual, 5.4\%--6.0\% for PTC first, and 2.5\%--4.8\% for PTC last [panel (a)]. Against five-step relaxation alone, the corresponding reductions are 62.5\%--74.2\%, 56.1\%--71.3\%, and 55.6\%--70.4\% [panel (b)].}

\revnov{PTC residual saves the most iterations but adds one Dirac application per iteration [Eq.~\eqref{eq:residual-insertion}]. PTC first [Eq.~\eqref{eq:pre}] reduces iterations without this extra cost. The placement with the fewest iterations therefore need not have the lowest work estimate in Eq.~\eqref{eq:count}.}

\begin{figure*}[!tp]
 \centering
\includegraphics[width=0.98\textwidth]{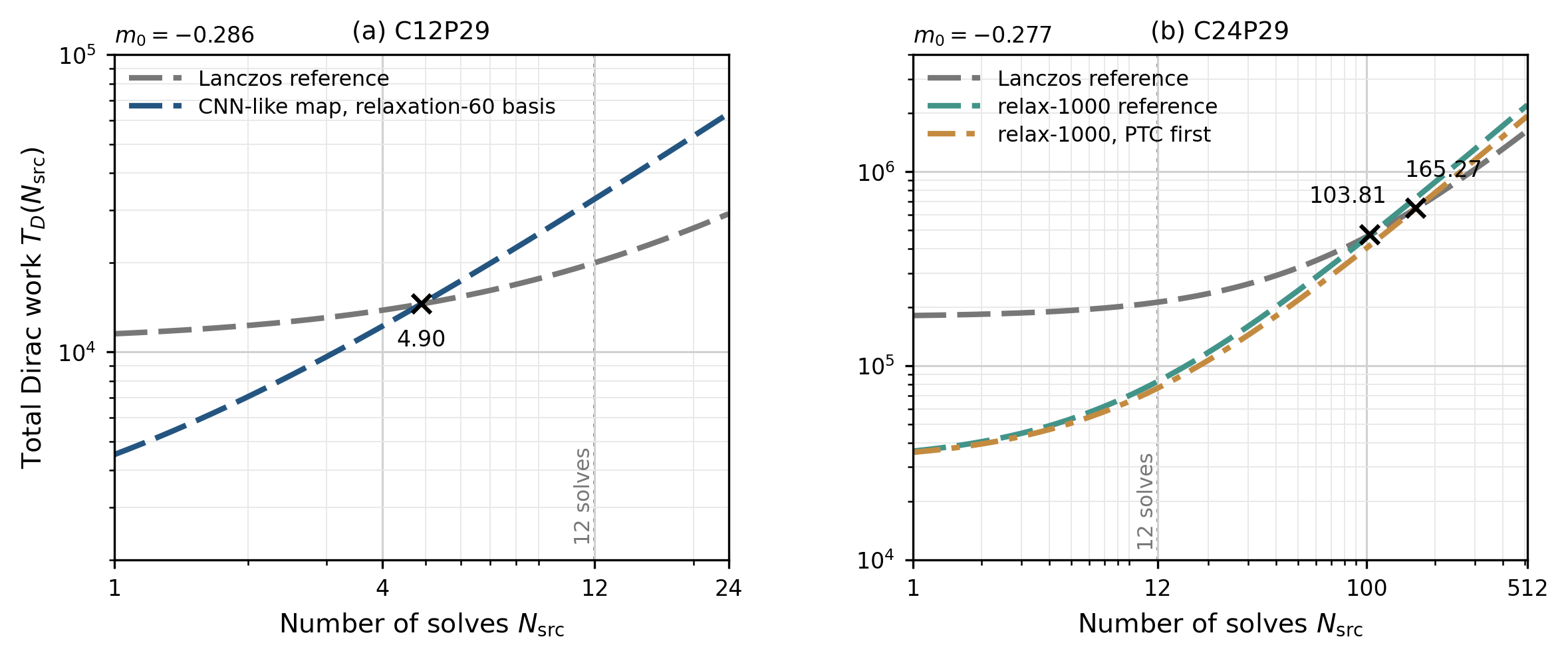}
\caption{Estimated total Dirac-operator application count from preparation and solve costs, according to Eq.~\eqref{eq:budget}, with logarithmic axes. Ensemble parameters are given in Table~\ref{tab:ensembles}. (a) C12P29 compares the Lanczos reference with the CNN-like map using candidate vectors after 60 relaxation steps. (b) C24P29 compares the Lanczos reference, the relaxation-1000 reference, and PTC first with the same relaxation basis. Vertical dotted lines mark 12 solves, corresponding to the twelve spin--color columns at one source position; crosses mark the estimated curve intersections at 4.90, 103.81, and 165.27 solves. Curves use representative costs per solve, and the intersections are extrapolated from those costs.}
\label{fig:source-budget}
\end{figure*}

\subsection{Preparation and solve costs across methods}
\label{subsec:amortization}
\revnov{Figure~\ref{fig:source-budget} combines preparation and solve costs using Eq.~\eqref{eq:budget}. On C12P29, the Lanczos reference and CNN-like map with 60-step relaxed candidates have preparation costs of 10742 and 1952 Dirac applications, and representative per-solve counts of 767 and 2561, respectively. The CNN-like preparation count includes the recorded projected-matrix assembly, although Eq.~\eqref{eq:learned-correction} does not require it. The estimated cost curves cross at 4.90 solves: the CNN-like method costs less for up to four solves, and the Lanczos reference costs less from the fifth onward.}

\revnov{On C24P29, PTC first [Eq.~\eqref{eq:pre}] and the relaxation-1000 reference share the same preparation cost. PTC reduces iterations without extra Dirac applications per iteration, lowering the estimated work per solve. The Lanczos reference costs more to prepare but less per solve. Its estimated cost crosses those of the relaxation reference and the PTC-first combination at 103.81 and 165.27 solves, respectively. These estimates assume available trained weights and fixed representative costs per solve; the preferred method depends on the number of sources.}

\section{Conclusion and outlook}
\label{sec:conclusion}
Deflation and multigrid exploit reduced representations of the slowly converging modes of the lattice Dirac operator. Our reference combines a projected template solve with fine-lattice relaxation. Within this framework, we studied two single-layer learned corrections for four-dimensional Wilson--Clover propagators. The gauge-covariant map $F_{\rm PTC}$ acts on fine-lattice quark fields with 144 shared complex parameters. The block map $F_{\rm CNN}$ uses 576 shared complex parameters to replace the reference solve for global template weights, followed by the same five Richardson steps. Both parameter counts are independent of lattice volume. The maps supply trial corrections to GCR, which retains the original Dirac equation and the true-residual criterion in Eq.~\eqref{eq:acceptance}.

We examined transfer without retraining, direct PTC fits to low-mode error, combinations of the two corrections, and the cost of preparation and repeated source solves. PTC transfers without retraining across the tested gauge backgrounds, lattice spacings, valence masses, and volumes. On the four A ensembles, transferred models give iteration reductions comparable to locally trained controls. On five held-out C24P29 configurations with twelve source spin--color columns each, the models trained on C12P29 and C24P29 reduce mean iterations by 31.5\% and 31.4\%, respectively, relative to unpreconditioned GCR. Direct PTC fits to low-mode error give little improvement, motivating its combination with a block correction. With the basis rebuilt on each target, fixed CNN-like weights reduce iterations by 42.1\%--51.1\% on C12P29 and by 60.5\% on C24P29 relative to five-step relaxation alone. The latter volume is 18 times the training volume; its iteration reduction relative to the reference correction with relaxation is 25.2\%.

Combining PTC with the CNN-like correction gives further gains. On C12P29, applying PTC to the residual left by the complete baseline reduces iterations by a further 15.2\%--19.7\%, averaged over four sources, but adds one Dirac application per iteration. Applying PTC first reduces iterations without that additional Dirac cost. Thus the placement that minimizes iterations need not minimize the estimated work. With trained weights available, preparation-plus-solve estimates favor inexpensive relaxed bases for fewer sources, while the lower per-source cost of a Lanczos reference can offset its higher preparation cost over many solves. On C24P29, PTC first lowers the solve cost of the relaxed-basis reference and extends its cost advantage over the Lanczos reference to more source solves. The useful comparison therefore includes both the target volume and the number of propagators required.

These results motivate incorporating the learned corrections into solvers with improved local bases and established multigrid implementations~\cite{Whyte2025Truncation,Whyte2025SVD,Frommer2014,Clark2016}. Wider-range corrections could be explored using generative models~\cite{KingmaWelling2014,Goodfellow2014} or attention-based models~\cite{Vaswani2017,Dosovitskiy2021}. Comparisons for production measurements should include training, basis preparation, local-map arithmetic, and communication alongside Dirac applications.

\begin{acknowledgments}
The work is supported in part by National Natural Science Foundation of China under grant No. 12375080, the Ministry of Science and Technology of China under Grant No. 2024YFA1611004, and by CUHK-Shenzhen under grant No. UDF01002851.
\end{acknowledgments}

\clearpage
\onecolumngrid
\appendix

\section{Quark fields and the propagator equation}
\label{app:sec:theory}

These appendices give the complete construction used in the quark-propagator solves, from candidate fields and local bases to the trial correction and its acceptance by the outer solver. The notation and training objectives are defined here independently of the main text.

Fix a gauge configuration $U$ and a dimensionless bare valence mass $m_0$. The Wilson--Clover Dirac operator is denoted by $D=D[U,m_0]$~\cite{SheikholeslamiWohlert1985}. For lattice extents $L_x,L_y,L_z,L_t$ in lattice units, the number of sites is $V=L_xL_yL_zL_t$. A quark field has four spin and three color components at every site, giving $n=4\times3\times V$ complex components. On C12P29, for example, $V=12^3\times32=55296$ and $n=4\times3\times55296=663552$. The Dirac matrix therefore has 663552 rows and columns. We use its sparse lattice action in an iterative solve.

For a source field $\eta$, the desired quark field $\psi$ satisfies
\begin{equation}
 D\psi=\eta.
 \label{app:eq:solve}
\end{equation}
A point source in one spin--color component gives one propagator column; twelve such solves give all source spin--color components at a fixed source position. The method below first prepares a representation of useful error corrections on the fixed gauge background. Each source solve then uses this representation to propose corrections, with generalized conjugate residual iteration (GCR)~\cite{Eisenstat1983} determining their contributions to $\psi$. We describe these operations in their execution order. Throughout, $\dagger$ denotes the Hermitian adjoint and $I_N$ the identity matrix on $N$ components.

\section{Preparation on a fixed gauge configuration}
\label{app:subsec:preparation}

\subsection{Generate candidate quark fields}
\label{app:subsec:prepared-fields}
The aim is to prepare fields that represent error components which are slow to relax. Let $N_v$ be the number of candidate fields, normally $N_v=32$. The inexpensive construction starts from independent complex Gaussian quark fields $v_i^{(0)}$, with $i=1,\ldots,N_v$. Applying $k$ steps of homogeneous relaxation with fixed step size $\omega_0$ gives
\begin{equation}
 v_i^{(k)}=(I_n-\omega_0D)^k v_i^{(0)}.
 \label{app:eq:candidates}
\end{equation}
  For this Richardson relaxation, we select the fixed step size $\omega_0$ with reference to the discussion of Richardson iteration in Ref.~\cite{Saad2003}. The principal inexpensive-template comparison uses $k=60$; other preparation lengths are specified with their numerical tests. The preparation step count $k$ is separate from the five relaxation steps applied during each propagator iteration.

For a more accurate low-mode preparation, we instead use Lanczos iteration~\cite{Lanczos1950} to obtain the eigenvectors associated with the smallest eigenvalues of $D^\dagger D$. These are the right singular vectors of $D$ with the smallest singular values. The normal operator acts on a quark field $v$ as $v\mapsto D^\dagger(Dv)$, without forming its matrix. It is used here to prepare the basis; the propagator equation remains Eq.~\eqref{app:eq:solve}. Collect the resulting candidates, from either construction, as the columns of $\mathcal V\in\mathbb C^{n\times N_v}$, where $\mathbb C$ denotes the complex numbers. Each column is a complete quark field.

\subsection{Construct the local block basis}
\label{app:subsec:local-basis}
Low-mode local coherence motivates representing these fields with a small basis on each lattice block~\cite{Luscher2007,Brannick2008,Babich2010}. Divide the lattice into nonoverlapping blocks of $2\times2\times2\times2=16$ sites. The block lattice has extents $(L_x/2,L_y/2,L_z/2,L_t/2)$ and contains $N_b=V/16$ blocks. Each block contains $4\times3\times16=192$ spin--color components.

On each block, restrict the first 16 candidate columns to its sites. Perform a QR factorization of the resulting $192\times16$ matrix, meaning a factorization into a matrix with orthonormal columns and an upper-triangular matrix; the implementation uses no column pivoting. Retain its first $s=8$ orthonormal columns as the local basis. When the first eight restricted candidates are locally independent, they determine this retained span; columns 9--16 do not enlarge it. A field in the resulting block representation has $s$ coefficients per block, hence
\begin{equation}
 n_c=N_b s
 \label{app:eq:coarse-dimension}
\end{equation}
coefficients in total. The subscript $c$ in $n_c$ labels the block representation rather than color. On C12P29, $N_b=55296/16=3456$ and $n_c=3456\times8=27648$.

Extend each retained local vector by zero outside its own block and collect these vectors as the columns of $P\in\mathbb C^{n\times n_c}$. Multiplication by $P$ reconstructs a quark field from its block coefficients. Its Hermitian adjoint $R=P^\dagger\in\mathbb C^{n_c\times n}$ extracts the coefficients of a fine-lattice field by projection. Local orthonormality gives $P^\dagger P=I_{n_c}$. Thus the first compression reduces 663552 fine-field components to 27648 block coefficients on C12P29. Both the reference and the learned block correction use this same local basis.

\subsection{Form the global templates for the reference correction}
\label{app:subsec:retained-space}
The reference further reduces the number of unknown correction coefficients. Project all candidate fields to the blocks and orthonormalize their independent coefficient patterns:
\begin{equation}
 W=\operatorname{orth}(R\mathcal V).
 \label{app:eq:global-basis}
\end{equation}
Here $\operatorname{orth}$ selects the numerical rank and orthonormalizes the retained columns. Let $N_d\leq N_v$ be their number, normally $N_d=32$. Then $W\in\mathbb C^{n_c\times N_d}$ contains $N_d$ global patterns of block coefficients. Reconstruct them on the fine lattice:
\begin{equation}
 Z=PW.
 \label{app:eq:selected-fields}
\end{equation}
The columns of $Z\in\mathbb C^{n\times N_d}$ are the template quark fields, with $Z^\dagger Z=I_{N_d}$. On C12P29 with $N_d=32$, $W$ is $27648\times32$ and $Z$ is $663552\times32$. Thus 32 counts the number of complete templates, each still containing 663552 components. A correction built from these templates is specified by only 32 global weights.

\subsection{Assemble the small Dirac matrix}
Apply $D$ to each template and project the result back onto the templates to obtain
\begin{equation}
 H=Z^\dagger DZ.
 \label{app:eq:retained-dirac}
\end{equation}
This $N_d\times N_d$ matrix describes the action of $D$ between the retained global templates. Its assembly requires $N_d$ applications of $D$. \revnov{With a prescribed real shift $\tau\geq0$, the matrix $H+\tau I_{N_d}$ is nonsingular in our tests. We factor it once and reuse the factorization for all source solves. Forming $W,Z$ and assembling this shifted matrix are needed by the reference correction.} The learned block replacement below acts directly in the basis $P$ and does not require this further compression to $N_d$ global directions when it is applied.

\subsection{Prepare the optional parallel-transport convolution (PTC) map}
\label{app:subsec:ptc-learning}
In the preprocessing placement, the fine-lattice learned operation acts on a quark field before the baseline correction is called. It combines the field at each site with its eight nearest neighbors, using PTC~\cite{LehnerWettig2023,LehnerWettigPooling2024}. Let $v$ be an input quark field and $x$ a fine-lattice site. Here $\delta$ denotes the zero displacement or one fine-lattice spacing in either direction along an axis.

Before combining the fields, the $3\times3$ color transporter $\mathcal U_\delta(x)\in\mathrm{SU}(3)$, where $\mathrm{SU}(3)$ is the special unitary group of degree three, brings the field at $x+\delta$ to $x$. A learned matrix $A_\delta\in\mathbb C^{4\times4}$ mixes its four spin components. The resulting map $F_{\rm PTC}$ is
\begin{equation}
 (F_{\rm PTC}v)(x)=\sum_{\delta}
 [A_\delta\otimes\mathcal U_\delta(x)]v(x+\delta),
 \label{app:eq:ptc}
\end{equation}
where $\otimes$ denotes the tensor product of the spin and color actions. The transporters use stout-smeared links~\cite{Morningstar2004}, with adjoint links for backward transport and $\mathcal U_0=I_3$ for the on-site term. Parallel transport ensures that $F_{\rm PTC}v$ transforms as a quark field under a gauge transformation.

The nine spin matrices are shared by all sites, giving $9\times4^2=144$ complex parameters. The color transporters are supplied by the gauge field. Training uses random probe fields $v$: the input is $Dv$, the target is $v$, and the spin matrices are fitted to make $F_{\rm PTC}Dv$ close to $v$. After training, applying $F_{\rm PTC}$ requires color transport and spin mixing but no application of $D$. These coefficients remain fixed throughout each propagator solve. The training objectives and numerical conventions are stated in Appendix~\ref{app:sec:training}.

\subsection{Prepare the optional convolutional neural network (CNN)-like map}
\label{app:subsec:coarse-learning}
The learned block operation uses the coefficients on each block and its eight nearest neighbors to propose a correction. Let $q\in\mathbb C^n$ be an input quark field. Its projection $Rq\in\mathbb C^{n_c}$ has $s$ coefficients $(Rq)(y)$ on block $y$. Let $\Delta$ denote the zero displacement or one block in either direction along each of the four axes. Assign one learned matrix $T_\Delta\in\mathbb C^{s\times s}$ to each of these nine displacements. The CNN-like map $F_{\rm CNN}$ is
\begin{equation}
 (F_{\rm CNN}Rq)(y)=\sum_{\Delta}T_\Delta(Rq)(y+\Delta).
 \label{app:eq:coarse-stencil}
\end{equation}
Each $8\times8$ matrix mixes the eight coefficients from one input block. Adding the nine contributions gives eight proposed correction coefficients on block $y$. All blocks share the same nine matrices, so this is a single CNN-like linear layer on the block lattice, with $9s^2=9\times8^2=576$ complex parameters independent of the lattice volume. The layer acts on coefficients of a target-built low-mode basis rather than directly on fine-lattice quark fields. Its output nevertheless has $n_c$ coefficients, one set of eight for every block.

To determine these matrices, we use residuals $q$ recorded at different iterations of earlier GCR solves on the training gauge field. Their projections $Rq$ are the training inputs. Target correction coefficients $\xi_c\in\mathbb C^{n_c}$ are obtained by an approximate GCR solve in the full block space:
\begin{equation}
 (RDP)\xi_c=Rq.
 \label{app:eq:coarse-system}
\end{equation}
The operator $RDP$ is $n_c\times n_c$, and $P\xi_c$ is the target fine-lattice correction. This is a projected Dirac solve in the full block basis $P$, whereas the reference uses the compressed templates $Z$ and $N_d$ global weights. A regularized complex least-squares fit chooses the nine matrices so that $F_{\rm CNN}(Rq)$ approximates $\xi_c$. These targets are prepared only for training; during a propagator solve, Eq.~\eqref{app:eq:coarse-stencil} supplies the block coefficients without a projected solve.

The block shifts are periodic, and training and application use the same local-basis convention. Different sources share the fitted matrices. They can also be reused on another gauge configuration or volume while rebuilding $P$ and $R$ for the target gauge field.

\section{A GCR iteration for each source}
\label{app:subsec:source-loop}

\subsection{Track the current residual}
For each new source, start from the zero initial approximation $\psi_0=0$. At iteration $j$, define the residual of the current approximation $\psi_j$ by
 \begin{equation}
 r_j=\eta-D\psi_j.
 \label{app:eq:true-residual}
\end{equation}
The unknown error is $e_j=\psi-\psi_j$. Subtracting the approximate equation from Eq.~\eqref{app:eq:solve} shows that it satisfies
\begin{equation}
 De_j=r_j.
 \label{app:eq:error-equation}
\end{equation}
Solving this equation exactly would itself require a full Dirac inversion. We instead use the prepared fields and local maps to obtain a trial correction. GCR updates $r_j$ by recurrence after each search step; Eq.~\eqref{app:eq:true-residual} defines that residual and is evaluated directly for the accuracy check. The following steps describe one call that supplies a trial correction to GCR.

\subsection{Choose the input, with or without PTC preprocessing}
Let $q\in\mathbb C^n$ denote the field passed into the baseline correction:
\begin{equation}
 q=
 \begin{cases}
 r_j, & \text{without PTC preprocessing},\\
 F_{\rm PTC}r_j, & \text{with PTC preprocessing}.
 \end{cases}
 \label{app:eq:baseline-input}
\end{equation}
All operations inside this call use $q$. In particular, when PTC preprocessing is enabled, both the first correction and the subsequent relaxation are driven by $F_{\rm PTC}r_j$. The residual of the original equation used by GCR remains $r_j$ from Eq.~\eqref{app:eq:true-residual}.

\subsection{Compute the first correction}
\label{app:subsec:first-correction}
\textit{Reference correction.}
\revnov{Project $q$ onto the global templates and solve for their weights $\xi\in\mathbb C^{N_d}$:
\begin{equation}
 (H+\tau I_{N_d})\xi=Z^\dagger q,
 \label{app:eq:amplitude-system}
\end{equation}}
using the prepared factorization, and reconstruct the first quark-field correction:
\revnov{\begin{equation}
 \delta\psi_c=Z\xi.
 \label{app:eq:coarse-action}
\end{equation}}
The subscript $c$ labels this template or block contribution. The field $\delta\psi_c$ has $n$ components; only its $N_d$ global weights were determined by the small solve. For $N_d=32$, this is the $32\times32$ solve in the reference method.

\revnov{Without the shift, the small solve imposes $Z^\dagger(q-D\delta\psi_c)=0$, the usual Galerkin condition. With the prescribed shift, this becomes $Z^\dagger(q-D\delta\psi_c)=\tau\xi$. The shift changes the small system, while the original propagator equation remains $D\psi=\eta$.}

\textit{Learned replacement.}
In this branch, project the input to $Rq$, apply $F_{\rm CNN}$ to obtain block-correction coefficients, and reconstruct the fine-lattice field:
\begin{equation}
 \delta\psi_c=PF_{\rm CNN}(Rq).
 \label{app:eq:learned-correction}
\end{equation}
This replaces the reference small solve and template reconstruction. It permits a separate set of correction coefficients on each block, determined from its local input, instead of restricting them to the global pattern \revnov{$W\xi$}. The two branches now proceed to the same relaxation step.

\subsection{Relax the remainder and return the trial field}
\label{app:subsec:relaxation}
  Apply relaxation to what remains after the first correction. Compute the remainder
\begin{equation}
 r_c=q-D\delta\psi_c.
 \label{app:eq:coarse-residual}
\end{equation}
Without PTC preprocessing, this is the residual of the still uncorrected error, $D(e_j-\delta\psi_c)=r_c$. With PTC preprocessing, it is the remainder of the auxiliary input $q=F_{\rm PTC}r_j$ inside the preconditioner.

Use five Richardson relaxation steps to construct a further correction. Let $u^{(h)}$ be the accumulated relaxation field after $h$ steps, and let $\omega$ be the fixed step size. Starting from $u^{(0)}=0$, iterate
\begin{equation}
 u^{(h+1)}=u^{(h)}+\omega(r_c-Du^{(h)}),
 \qquad h=0,\ldots,4.
 \label{app:eq:smoother}
\end{equation}
Denote the resulting five-step map by $S$, so that $Sr_c=u^{(5)}$. The trial field $t_j$ returned to GCR is the sum of the first correction and the relaxation correction:
\begin{equation}
 t_j=\delta\psi_c+Sr_c.
 \label{app:eq:trial-correction}
\end{equation}
This order lets the prepared representation supply one part of the correction and relaxation address the remaining input with inexpensive applications of $D$.

\revnov{Write $Bq=t_j$ for the complete correction, whether its first step uses the global templates or the CNN-like map $F_{\rm CNN}$. For relaxation alone, $B=S$ acts directly on $q$. Without PTC preprocessing, $t_j=Br_j$; for PTC first,
\begin{equation}
 t_j=B(F_{\rm PTC}r_j).
 \label{app:eq:pre}
\end{equation}
Standalone PTC returns $t_j=F_{\rm PTC}r_j$, without a block correction or relaxation.} Section~\ref{app:subsec:composition} defines the other placements of $F_{\rm PTC}$.

\subsection{Update the propagator and test the original equation}
GCR computes $Dt_j$ and orthogonalizes it against the Dirac images of the retained search directions in the current restart cycle. Applying the same linear combinations to $t_j$ gives the search direction $p_j$. Its Dirac image $Dp_j$ is therefore orthogonal to those retained images. Let $\|\cdot\|_2$ denote the Euclidean norm over all field components. For nonzero $Dp_j$, GCR minimizes the residual along $p_j$ by choosing the step coefficient $\alpha_j$:
\begin{equation}
 \alpha_j=\frac{(Dp_j)^\dagger r_j}{\|Dp_j\|_2^2}.
 \label{app:eq:gcr-coefficient}
\end{equation}
The propagator approximation is then updated by
\begin{equation}
 \psi_{j+1}=\psi_j+\alpha_j p_j.
 \label{app:eq:gcr-update}
\end{equation}
The corresponding residual is $r_{j+1}=r_j-\alpha_j Dp_j$. Thus GCR chooses the contribution of the proposed field using the original residual $r_j$, including when the preconditioner was fed $F_{\rm PTC}r_j$.

\revnov{We recompute the true residual from the original Dirac action and accept a source solve at the tolerance used in the numerical tests:}
\revnov{\begin{equation}
 \frac{\|\eta-D\psi_j\|_2}{\|\eta\|_2}\leq10^{-8}.
 \label{app:eq:acceptance}
\end{equation}}
\revnov{If this test fails, the iteration continues from the updated residual.} The next source reuses the preparation and repeats this solve and accuracy test.

\section{What the two learned operations change}
\label{app:sec:learned-operations}
The single-layer, gauge-covariant PTC~\cite{LehnerWettig2023,LehnerWettigPooling2024} has 144 shared complex parameters: nine shared $4\times4$ spin matrices, one for each of the on-site and eight nearest-neighbor paths; the gauge links supply color transport. We first test it alone in the outer GCR solve: at each iteration, it maps the current residual $r_j$ to the trial correction $F_{\rm PTC}r_j$, with no block correction or relaxation, and we assess its effect on the iteration count. We then combine PTC with the multigrid-inspired correction described above, comparing placements before and after that correction and on the residual it leaves. These placements can affect outer GCR iterations and counted Dirac applications differently. In all cases, PTC changes the trial correction supplied to GCR; GCR continues to solve the original equation and uses the same true-residual test.

The CNN-like map acts instead on coefficients in the local block basis. Its nine shared $s\times s$ matrices keep the learned parameter count at $9s^2$ as the volume grows. Candidate-field preparation, the local basis, the projected Dirac equation, and relaxation follow established deflation and multigrid ideas. In the reference sequence, the residual is projected onto global templates, their weights are determined by the projected Dirac system, and the resulting correction is applied before relaxing the remainder. The CNN-like map replaces this first correction: it uses the same local basis but computes a correction on each block from the coefficients on that block and its neighbors, rather than solving for global weights. Within the multigrid-inspired sequence, the map supplies the first correction, after which the same relaxation acts on the remaining residual and the resulting trial field goes to GCR. The preparation and source-by-source iteration are specified in Appendices~\ref{app:subsec:preparation} and \ref{app:subsec:source-loop}. The required preparation is performed once for a fixed gauge field and valence mass and reused across sources. Both branches use the same five Richardson steps and true-residual test. For the relaxation-only baseline, omit basis preparation and the first correction and apply the five relaxation steps directly to $q$. Standalone PTC instead returns $F_{\rm PTC}r_j$ to GCR without a block basis or relaxation. Between direct checks of the original equation, GCR updates the residual by recurrence.

\section{Training definitions and numerical conventions}
\label{app:sec:training}

\subsection{PTC and block-map objectives}
The PTC training uses $N_p$ complex random probe fields $v_\ell$, with probe index $\ell=1,\ldots,N_p$, in each optimization step. The input--target pair is $(Dv_\ell,v_\ell)$, and the normalized loss $\mathcal L_{\rm PTC}$ is
\begin{equation}
 \mathcal L_{\rm PTC}=\frac{1}{N_p}\sum_{\ell=1}^{N_p}
 \frac{\|F_{\rm PTC}Dv_\ell-v_\ell\|_2^2}{\|v_\ell\|_2^2}.
 \label{app:eq:ptc-loss}
\end{equation}
The transporters use one stout-smearing step with parameter $\rho=0.125$. The transfer models are trained for 900 steps with adaptive moment estimation (Adam) and learning rate $10^{-3}$. After training, only the target gauge links change when these weights are transferred.

For the block map, the 64 training pairs are $(Rq_\ell,\xi_{c,\ell})$, where $\ell$ indexes the pairs, $q_\ell$ is a residual recorded in an earlier GCR solve, and $\xi_{c,\ell}$ is an approximate solution of Eq.~\eqref{app:eq:coarse-system}. The regularized least-squares loss is
\begin{equation}
 \mathcal L_{\rm CNN}=\sum_{\ell=1}^{64}\|F_{\rm CNN}(Rq_\ell)-\xi_{c,\ell}\|_2^2
 +\lambda\sum_\Delta\|T_\Delta\|_F^2.
 \label{app:eq:coarse-loss}
\end{equation}
Here $\lambda\geq0$ is the regularization strength, and the Frobenius norm $\|\cdot\|_F$ is the square root of the sum of squared matrix-entry magnitudes. The penalty suppresses large entries of the shared block matrices. Training and application use the same local-basis convention. The learned matrices remain fixed, while the local maps $P$ and $R$ are reconstructed for each target configuration.

\subsection{Basis regularization and source solves}
The global orthonormalization uses a relative numerical-rank threshold of $10^{-8}$, normally retaining $N_d=N_v=32$ templates. Candidate preparation applies no intermediate normalization or global reorthogonalization. The relaxation step sizes obey $\omega=\omega_0$. \revnov{The real shift in Eq.~\eqref{app:eq:amplitude-system} is
\begin{equation}
 \tau=10^{-8}\bigl(|\operatorname{tr}H|/N_d+10^{-300}\bigr),
 \label{app:eq:shift-value}
\end{equation}}
where $\operatorname{tr}$ is the matrix trace. Each source solve starts from zero, uses GCR restart length 50 and two orthogonalization passes, and accepts only a true relative residual \revnov{at most $10^{-8}$}.

\needspace{8\baselineskip}
\subsection{Alternative placements of PTC}
\label{app:subsec:composition}
\revnov{Let $B$ denote the complete baseline correction, including the five Richardson steps when a template or block correction is used. PTC first returns $B(F_{\rm PTC}r_j)$, as in Eq.~\eqref{app:eq:pre}. PTC last applies $F_{\rm PTC}$ after the baseline:
\begin{equation}
 t_j=F_{\rm PTC}(Br_j).
 \label{app:eq:post}
\end{equation}
For PTC residual, apply $F_{\rm PTC}$ to the residual left after the complete baseline correction:
\begin{equation}
 t_j=Br_j+F_{\rm PTC}\bigl(r_j-D(Br_j)\bigr).
 \label{app:eq:residual-insertion}
\end{equation}
Here $r_j-D(Br_j)$ is formed after the complete baseline, including relaxation; $r_c$ in Eq.~\eqref{app:eq:coarse-residual} is formed after only the first template or block correction. Computing $D(Br_j)$ requires one additional Dirac application.} All three placements supply a trial field to the same outer GCR iteration and use the acceptance test in Eq.~\eqref{app:eq:acceptance}.

\twocolumngrid
\bibliography{ref}

\end{document}